\documentclass[aps,pra,reprint,superscriptaddress,amsmath,amsfonts,amssymb]{revtex4-1}

\usepackage{graphicx}
\usepackage{hyperref}
\usepackage{amsthm}
\usepackage{physics}
\usepackage[english]{babel}
\usepackage{qcircuit}
\usepackage{xspace}

\newcommand{\uba}{Universidad de Buenos Aires, Facultad de Ciencias Exactas y Naturales, Departamento de Física, Buenos Aires, Argentina}
\newcommand{\ifiba}{CONICET - Universidad de Buenos Aires, Instituto de Física de Buenos Aires (IFIBA), Buenos Aires, Argentina}
\newcommand{\oxford}{Department of Physics and Astronomy, University of Exeter, Stocker Road, Exeter EX4 4QL, UK}
\newcommand{\exeter}{Department of Engineering Science, University of Oxford, Parks Road, Oxford OX1 3PJ, UK}

\newcommand{\Id}{\mathbb{I}}

\newcommand{\TPM}{TPM\xspace}
\newcommand{\OPM}{SM\xspace}

\newcommand{\adj}[1]{#1^{\dagger}}

\newcommand{\mean}{\expval}

\newcommand{\opmark}[1]{{\hat{#1}}}

\newcommand{\dephased}[1]{\bar{#1}}

\NewDocumentCommand{\DiracDelta}{}{\opbraces{\delta}}

\newcommand{\ExpInline}[1]{\exp\left[#1\right]}
\newcommand{\ExpSuper}[1]{{\mathrm{e}}^{#1}}

\newcommand{\Normal}{\mathcal{N}}
\newcommand{\NormalDist}[3]{\Normal\left(#3\,\middle|\,#1,#2\right)}

\newcommand{\Etilde}{\tilde{E}}

\newcommand{\Htilde}{\tilde{H}}

\newcommand{\Pitilde}{\tilde{\Pi}}

\newcommand{\quasiPtpmGaussName}{P_{\mathcal{N}}}
\newcommand{\quasiPtpmGauss}[2]{\quasiPtpmGaussName\left(#1\middle|#2\right)}
\newcommand{\quasiw}{w}
\newcommand{\tpmHini}{H}
\newcommand{\tpmHfin}{\Htilde}

\newcommand{\tpmEini}{E}
\newcommand{\tpmEfin}{\Etilde}

\newcommand{\tpmUdriving}{\mathcal{U}}

\newcommand{\tpmw}{w}

\newcommand{\tpmSys}{\mathcal{S}}

\newcommand{\tpmProjEini}{\Pi}
\newcommand{\tpmProjEfin}{\Pitilde}

\newcommand{\tpmPtrans}[2]{{p}_{{#2}|{#1}}}
\newcommand{\tpmPmeas}[1]{{p}_{{#1}}}
\newcommand{\tpmPwork}{P_\mathrm{\TPM}}

\newcommand{\IntReals}[1]{\int_{-\infty}^{\infty}\dd{#1}}

\newcommand{\quasiPwork}{P_W}
\newcommand{\quasiVarw}{w}
\newcommand{\quasiVart}{\tau}
\newcommand{\quasiVarInt}{y}

\newcommand{\PhaseSpaceInt}{%
  \int_{-\infty}^{\infty}\dd{\quasiVart}\int_{-\infty}^{\infty}\dd{\quasiVarw}}

\newcommand{\wpovmSys}{\tpmSys}
\newcommand{\wpovmAux}{\mathcal{A}}

\newcommand{\wpovmAuxQop}{\mathcal{W}_{\wpovmAux}}
\newcommand{\wpovmAuxPop}{\mathcal{T}_{\wpovmAux}}

\newcommand{\wpovmAuxPeig}{p}

\newcommand{\wpovmGaussianState}[2]{{#1},{#2}}

\newcommand{\quasiSys}{\wpovmSys}
\newcommand{\quasiAux}{\wpovmAux}

\newcommand{\quasiIniSys}{\rho_{\wpovmSys}}
\newcommand{\quasiIniAux}{\rho_{\wpovmAux}}
\newcommand{\quasiFinAux}{\rho_{\wpovmAux}(t_f)}
\newcommand{\quasiIniSysTevol}[1]{\quasiIniSys\left(#1\right)}
\newcommand{\quasiIniSysDephased}{\dephased{\rho}_{\wpovmSys}}

\newcommand{\quasiHini}{\tpmHini}
\newcommand{\quasiHfin}{\tpmHfin}

\newcommand{\quasiEini}{\tpmEini}
\newcommand{\quasiEfin}{\tpmEfin}

\newcommand{\quasiProjEini}{\tpmProjEini}
\newcommand{\quasiProjEfin}{\tpmProjEfin}

\newcommand{\quasiUdriving}{\mathcal{U}}

\newcommand{\quasiGaussianStd}{\sigma}
\newcommand{\quasiGaussianStdWT}{{\sigma}}

\newcommand{\tpmRhoIni}{\rho_\mathcal{S}}
\newcommand{\dtpmRhoIni}{\dephased{\rho}_\mathcal{S}}

\usepackage{xcolor}

\begin{document}

\title{A Wigner quasiprobability distribution of work}
\author{Federico Cerisola}
 \email{federico.cerisola@eng.ox.ac.uk}
 \affiliation{\uba} \affiliation{\ifiba} \affiliation{\oxford} \affiliation{\exeter}
\author{Franco Mayo}
  \affiliation{\uba} \affiliation{\ifiba}
\author{Augusto J. Roncaglia}
  \affiliation{\uba} \affiliation{\ifiba}


\begin{abstract}
In this article we introduce a quasiprobability distribution of work that is based on the Wigner function. This construction rests on the idea that the work done on an isolated system can be coherently measured by coupling the system to a quantum measurement apparatus.  In this way, a quasiprobability distribution of work can be defined in terms of the Wigner function of the apparatus. This quasidistribution contains the information of the work statistics and also holds a clear operational definition. Moreover, it is shown that the presence of quantum coherence in the energy eigenbasis is related with the appearance of characteristics related to non-classicality in the Wigner function such as negativity and interference fringes. On the other hand, from this quasiprobability distribution it is straightforward to obtain the standard two-point measurement probability distribution of work and also the difference in average energy for initial states with coherences. 
\end{abstract}

\maketitle
\section{Introduction}
The notion of work is one of the most basic and fundamental concepts in physics, particularly in thermodynamics. 
During the last decades, several attempts have been made to obtain the work statistics for non-equilibrium thermodynamic transformations in the quantum regime. These definitions where motivated by the idea of extending classical fluctuation theorems~\cite{jarzynski1997, crooks1999, talkner2007, talkner2007a, campisi2011} to the quantum regime.
 In order to describe the thermodynamics of general non-equilibrium quantum processes, it is necessary to provide a general definition of work valid for any quantum system and process. However, this task presents serious difficulties. This is due, in the first place, to the fact that many concepts belonging to the classical definition of work cannot be translated to quantum mechanics. For example, the basic definition of the work that a force performs on a particle along a trajectory cannot be used in quantum mechanics because of the lack of meaning of trajectories in the theory, although recently a definition of quantum work was made by considering bohmian trajectories~\cite{sampaio2018quantum}. 
A great advance came in the area with the definition of the two point measurement protocol~(\TPM) to define work in driven isolated quantum systems~\cite{tasaki2000,kurchan2000,talkner2007,talkner2007a}. This definition is based on the simple observation that, for an isolated system,  work is a random variable associated to the difference in energy along the process. Thus, in order to determine this random value one should make an energy measurement at the beginning and another at the end of the process. This definition is not only straightforward in an operational sense, but it also recovers the results of the fluctuation theorems for quantum systems~\cite{tasaki2000,kurchan2000,talkner2007,talkner2007a,esposito2009,campisi2011} and was verified experimentally in different platforms~\cite{batalhao2014, an2015,  cerisola2017,smith2018,hernandez2020experimental, solfanelli2021experimental}. 

However, there is a caveat with the \TPM  when one considers initial states that have coherences in the energy basis. This is because the first energy measurement destroys these coherences, and therefore the \TPM scheme is insensitive to quantum coherence between different energy subspaces. This leads to undesirable consequences, for instance, related with the fact that the average work done in the process is different from the change in the average energy of the system. Moreover, it has been shown that it is impossible to define a probability distribution of work that satisfies at the same time, the fluctuation theorems and whose mean value of work equals the average energy change for states with coherences~\cite{perarnau-llobet2017}. This has led the community to consider different approaches to generalize the work distribution~\cite{lostaglio2018b,xu2018,sagawa2012,sampaio2018quantum} including some proposals for quasiprobability distributions of work~\cite{allahverdyan2014,solinas2015,solinas2016,wiseman2002,hall2004,miller2017, lobejko2022work,francica2022class, francica2022most}. 

In this article, we propose a distribution based on the Wigner function. This construction relays on the fact that the work probability distribution can also be coherently measured by  coupling the system to a quantum  apparatus and making a single measurement over the apparatus, i.e. a single-measurement protocol (\OPM)~\cite{roncaglia2014,de-chiara2015,cerisola2017}. In this way, the final state of the apparatus contains the information about the work distribution and one can define a quasiprobability distribution~\cite{cerisolathesisp2020}. This approach  provides a clear operational definition with an immediate experimental implementation. In addition, the Wigner function is represented using coordinates that have an intuitive interpretation in terms of time and energy associated with the work. Moreover, it can be shown that the presence of quantum coherence is related with the appearance of characteristics related to non-classicality in the Wigner function, such as negativity and interference fringes. On the other hand, for coherence free states this definition agrees with the standard two point measurement probability distribution of work. 

The paper is organized as follows. In Sec.~\ref{sec:workdistribcoh:tpmcoh}  we briefly discuss the two point measurement scheme, and the single measurement protocol. In Sec.~\ref{sec:workdistribcoh:workquasi:work} we introduce the  quasiprobability distribution of work based on the Wigner function, showing how it works for initial states of the system with and without coherence. In Sec.~\ref{sec:workdistribcoh:workquasi:experimental} we discuss experimental implementations, and we end with discussions and conclusion in Sec.~\ref{section: conclusions}. 

\section{Work Statistics}
\label{sec:workdistribcoh:tpmcoh}

As mentioned, we are interested in the work distribution for isolated quantum systems that are subjected to an external driving.  In this way, the external work can be associated to the energy change of the system. The typical scenario consists of a system $\quasiSys$ that starts in a given initial state, $\tpmRhoIni$, and after that, a driving represented by a unitary evolution $\tpmUdriving$ is applied. The driving is such that it changes the Hamiltonian from an initial $\tpmHini$ to a final one $\tpmHfin$
\begin{equation}
  \quasiHini = \sum_{n} \quasiEini_n \quasiProjEini_n,
  \qquad
  \quasiHfin = \sum_{m} \quasiEfin_m \quasiProjEfin_m,
\end{equation}
where  $\tpmProjEini_n$ are the projectors on each energy subspace.
In this case, what we know is that the average change of energy in the system is
\begin{equation}
  \Delta E =
 \tr\left[\tpmHfin\,\tpmUdriving\,\tpmRhoIni\,\adj{\tpmUdriving}\right] -
  \tr\left[\tpmHini\,\tpmRhoIni\right],
  \label{eq:cohEnergyDiff}
\end{equation}
where  $\tpmUdriving\,\tpmRhoIni\,\adj{\tpmUdriving}$ is the final state of the system.
Clearly, it would be desirable that the average work obtained from the corresponding probability distribution equals this average energy change. This requisite is equivalent to asking that the first law of thermodynamics for mean values is satisfied for an isolated system. However, it can be shown that if one imposes that the statistics of work is consistent with the standard fluctuation theorems, the distribution of work should be defined by the two-point measurement protocol~\cite{tasaki2000,kurchan2000,talkner2007,talkner2007a}. In this case, although the resulting work average coincides with the mean energy difference for initial equilibrium states (diagonal in the initial energy eigenbasis), it is different for initial states with coherences.

\subsection{The two-point measurement protocol}

The two-point measurement protocol allows us to define a probability distribution of work consistent with fluctuation theorems. 
In order to do that, one should define a work value for each realization. This is done in terms of the difference of two energy values that are obtained by making two energy measurements, one at the beginning and the other one at the end of the driving. In this way, the corresponding probability distribution can be written as
 \begin{equation}
     P_\mathrm{\TPM}(w) = \sum_{n,m} p_n p_{m\vert n} \ \delta\left(w - (\Tilde{E}_m - E_n)\right),
     \label{eq:def:wpovm:probwnm}
 \end{equation}
where $p_n$ is the probability of obtain $E_n$ in the first energy measurement, and $p_{m\vert n}$ is the conditional probability of obtaining $\tilde E_m$ at the end given that $E_n$ was obtained at the beginning. Therefore, if the initial state is already diagonal in the energy eigenbasis, the first measurement does not modify the state, and it can be easily verified that the mean value of work equals the average energy difference.
Indeed, from this probability distribution one can evaluate the mean value of work as
\begin{align}
  \expval{\tpmw}
  &= \int\dd{w} \tpmPwork(\tpmw)\ \tpmw
  = \sum_{n,m} \tpmPmeas{n} \tpmPtrans{n}{m}
    \left(\tpmEfin_m - \tpmEini_n\right) \nonumber \\
  &= \sum_{n,m} \tr\left[ \tpmProjEfin_m \,\tpmUdriving \,\tpmProjEini_n \,\tpmRhoIni\,
     \tpmProjEini_n \,\adj{\tpmUdriving} \right]
     \left(\tpmEfin_m - \tpmEini_n\right) \nonumber \\
  &= \tr\left[\tpmHfin \,\tpmUdriving \,\dtpmRhoIni\,
    \adj{\tpmUdriving}\right] - \tr\left[\tpmHini\,\dtpmRhoIni\right],
    \label{eq:meanw}
\end{align}

where $\dtpmRhoIni = \sum_{n} \tpmProjEini_n \tpmRhoIni
\tpmProjEini_n$ is the dephased initial state. This state is obtained by removing all the coherences between different energy subspaces of the initial Hamiltonian, and it is equivalent to the state resulting the following asymptotic temporal average 
\begin{equation}
  \dtpmRhoIni = \lim_{T\to\infty} \frac{1}{T} \int_{-T/2}^{T/2}\dd{t}
    e^{-\frac{i}{\hbar}\tpmHini t} \tpmRhoIni e^{\frac{i}{\hbar}\tpmHini t}.
    \label{eq:timeaverage}
\end{equation}
Therefore, unless the initial state is diagonal in the basis of the initial hamiltonian, the average given by the \TPM is different from the average energy difference of the system. 
In fact, if the initial state is diagonal in this basis, then $\rho_\mathcal{S}$ can be interpreted as a 
`classical' probability distribution over the different energies. In that case, the measurement is not invasive, in the sense that it only reveals the value of the energy in each realization of the experiment. On the other hand, for an initial state with coherences, the initial energy is not well defined and this interpretation is not straightforward.

\subsection{The single-measurement protocol}

Another method for assessing to the work probability distribution was introduced
in~\cite{roncaglia2014}. The method is based on the idea that the work measurement can be described in terms of a generalized measurement (POVM). That is, by coupling the system to an ancilla that is finally subjected to a `single measurement' (SM). In this way, it can be shown that one can obtain the same probability distribution provided the ancilla is properly initialized. 

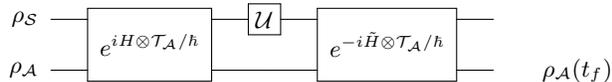
\begin{figure}
\begin{equation*}
\Qcircuit @C=1.5em @R=1em {
\lstick{\tpmRhoIni} & \multigate{1}{e^{i \tpmHini\otimes \mathcal{T}_\mathcal{A}/\hbar}} & \gate{\mathcal{U}} & \multigate{1}{e^{-i \tpmHfin\otimes  \mathcal{T}_\mathcal{A}/\hbar}} & \qw & \rstick{}  \\
\lstick{\quasiIniAux} & \ghost{e^{i \tpmHini\otimes \mathcal{T}_\mathcal{A}/\hbar}} & \qw & \ghost{e^{-i \tpmHfin\otimes \mathcal{T}_\mathcal{A}/\hbar}} & \qw & \rstick{\quasiFinAux} 
}
\end{equation*}
\caption{Circuit that describes the single-measurement protocol from where the work probability distribution can be obtained.}
    \label{fig:circuit}
\end{figure}

Let us now describe briefly the general method that is summarised in the circuit of Fig.~\ref{fig:circuit}. Initially, the system is in the state $\quasiIniSys$ and there is an auxiliary system (ancilla) $\quasiAux$ whose state is described terms of a continuous degree of freedom. 
In the ancilla's space one can consider two canonically conjugated operators, $\wpovmAuxQop$ and $\wpovmAuxPop$, such that $[\wpovmAuxQop,\wpovmAuxPop]=i \hbar$. 
Thus, the evolution contains two coherent interactions between $\quasiSys$ and $\quasiAux$: one at the beginning $\ExpSuper{i\quasiHini\otimes\wpovmAuxPop/\hbar}$ and another $\ExpSuper{-i \quasiHfin\otimes\wpovmAuxPop/\hbar}$ at the end of the driving. 
Thus, each interaction can be viewed either as a coherent translation in the variable $w$ of the ancilla in an amount that depends on the energy of the system or, conversely, as a coherent time-translation (free evolution) of the system whose time interval is proportional to the variable $\tau$ of the ancilla. Therefore, one can immediately associate the variables $w$ and $\tau$ to energy (work) and time, respectively. 

Following the protocol of the circuit, if the initial state of $\quasiAux$ is $\quasiIniAux$, after the interaction with the system its reduced state is
\begin{align}
  \quasiFinAux = \sum_{n,n',m}
    \tr\left[\quasiProjEfin_{m} \quasiUdriving \quasiProjEini_n \quasiIniSys
    \quasiProjEini_{n'} \adj{\quasiUdriving} \right] & \nonumber \\
    \ExpSuper{-i\quasiw_{nm}\wpovmAuxPop/\hbar}
    \quasiIniAux
    \ExpSuper{i\quasiw_{n'm}\wpovmAuxPop/\hbar},
\end{align}
where $\quasiw_{nm} = \quasiEfin_m - \quasiEini_n$ are the different work values.
The \OPM protocol finishes by performing a projective measurement of the observable $\wpovmAuxQop$. In this case, for highly localized initial pure states of $\quasiAux$, the resulting probability distribution is equivalent to the work distribution of the \TPM protocol. 
Notably, within this formulation one can associate work to an observable acting over the ancillary system. Of course, work is not an observable acting on the system's space~\cite{talkner2007a}.

It is important to stress at this point that the entangling interaction between system and apparatus establishes a coherent record of the different values of work. Therefore, the reduced state of the ancilla contains information not only about the probability distribution given by the \TPM, but also about the initial state of the system. 
At the end, the type of measurement that is done over the ancilla (or the battery), determines which information is extracted from the protocol.
It is also interesting to notice that this type of interaction appears in a very related task: the work extraction from a quantum system. This can be modeled by adding an interaction between the system and an auxiliary system that acts as a battery in which work is stored~\cite{skrzypczyk2014, alhambra2016fluctuating, richens2016}. In general, the battery can be thought of as a continuous variable system, an ideal weight, with a hamiltonian like the operator $\wpovmAuxQop$. The work extraction process consists on some unitary evolution on the joint system (where the driving on the system is included) that can change the system hamiltonian from $\tpmHini$ to $\tpmHfin$. The extracted work, in this way, is stored in the battery. There are a few conditions that should be imposed in this framework in order to ensure that the weight does not provide any thermodynamical resource to the work extraction process~\cite{alhambra2016fluctuating}, one of them is of course energy conservation.
It has been shown in~\cite{alhambra2016fluctuating} that the unitary operations that satisfy these conditions are of the form $e^{i\tpmHini\otimes\wpovmAuxPop/\hbar}\left(\quasiUdriving\otimes\mathbb{I}_\quasiAux\right)e^{-i\tpmHfin\otimes\wpovmAuxPop/\hbar}$
where $\quasiUdriving$ is the driving of the system. Therefore, it is straightforward to see that these are the same operations (up to a sign) used in the \OPM protocol for measuring work. Thus, there is also a clear operational interpretation of the state of the ancilla as the state of a battery where work is stored.\\

\section{The Wigner  distribution of work}
\label{sec:workdistribcoh:workquasi:work}

In the following, we will define a generalized work distribution. The general idea is inspired by the SM protocol. As we just mentioned, the state of the ancilla after the interaction not only holds information about work, but also about the coherences present in the initial state. In order to extract such information, we will evaluate their Wigner function~\cite{wigner1932,hillery1984},
$\quasiPwork$. The Wigner function is a quasi-probability distribution that is used to represent quantum states in phase space. This is a real-valued function that, unlike their classical counterparts, can be negative for generic quantum states. This property has been widely used  as an indicator of quantumness in different contexts, for instance in the study of the quantum-classical transition. 

In our case, we will define it for the final state of the ancilla and in terms of the conjugate variables $\quasiVarw$ and $\quasiVart$ as
\begin{widetext}
\begin{align}
  \quasiPwork(\quasiVarw, \quasiVart)
  &= \frac{1}{2\pi\hbar} \IntReals{\quasiVarInt}
    \matrixel{\quasiVarw + \frac{\quasiVarInt}{2}}
      {\quasiFinAux}{\quasiVarw - \frac{\quasiVarInt}{2}}
    \ExpSuper{-i\quasiVart\quasiVarInt/ \hbar}
    \nonumber \\
  &= \frac{1}{2\pi\hbar} \sum_{n,n',m} \tr\left[ \quasiProjEfin_{m}
    \quasiUdriving \quasiProjEini_n \quasiIniSys \quasiProjEini_{n'}
    \adj{\quasiUdriving} \right]   \IntReals{\quasiVarInt} 
    \matrixel{\quasiVarw + \frac{\quasiVarInt}{2} -
        \quasiw_{nm}}
      {\quasiIniAux}{\quasiVarw - \frac{\quasiVarInt}{2} -
        \quasiw_{n'm}}
    \ExpSuper{-i\quasiVart\quasiVarInt/ \hbar}
  \label{eq:def:wquasi:wwigner:general}
\end{align}
\end{widetext}
This  expression is valid for a generic initial state of the ancilla.
In order to evaluate it, we will assume that the initial state of the ancilla is a coherent Gaussian state. This assumption not only will allow us to easily perform analytical calculations, but is also and appropriate choice for the description of typical experimental situations. Moreover, Gaussian states are classical, in the sense that they have a positive Wigner function. This guarantees that any negativity appearing in the Wigner function of the ancilla comes exclusively from their interaction with the system.  
Thus, we consider $\quasiIniAux = \dyad{\wpovmGaussianState{0}{\sigma}}$ a coherent Gaussian state with zero mean and variance $\quasiGaussianStd^2$ in $\wpovmAuxQop$ (and hence zero mean and variance $\frac{\hbar^2}{2\quasiGaussianStd^2}$ in $\wpovmAuxPop$). After replacing this in Eq.~\eqref{eq:def:wquasi:wwigner:general} (see Appendix \ref{sec:appQuasi}) and using that $\quasiProjEini_n \quasiIniSys \ExpSuper{i\quasiVart \quasiEini_{n} /\hbar}=\quasiProjEini_n \ExpSuper{i\quasiVart   \quasiHini/\hbar} \quasiIniSys$ we obtain an expression for the quasiditribution of work for a generic process
 \begin{eqnarray}    
  && \quasiPwork(\quasiVarw, \quasiVart)
  = \sum_{n,n',m} \tr\left[ \quasiProjEfin_{m}
    \quasiUdriving \quasiProjEini_n \quasiIniSys(-\quasiVart) \quasiProjEini_{n'}
    \adj{\quasiUdriving} \right] \nonumber \\
  &&  \times \ \NormalDist{\frac{\quasiw_{nm} + \quasiw_{n'm}}{2}} 
      {\quasiGaussianStdWT}{\quasiVarw} 
     \NormalDist{0}{\frac{\hbar}{\sqrt 2\quasiGaussianStdWT}}{\quasiVart}, 
  \label{eq:def:wquasi:gaussian}
\end{eqnarray}
where $\NormalDist{\mu}{\sigma}{\quasiVarw}$ is a normal probability density in $\quasiVarw$ with mean value $\mu$ and variance $\sigma^2$ (analogously for $\quasiVart$).
From this expression we can appreciate again the operational interpretation of the variables $\quasiVarw$ and $\quasiVart$ that characterize the state of the ancilla.

In the following, we will introduce some notation that will be useful to simplify forthcoming expressions.
First, let us  recall that the distribution $ \tpmPwork(\quasiw)$ does not take into account any coherence between the different energy subspaces  of $H$ in the initial state $\quasiIniSys$. Therefore, we can associate this probability distribution to the dephased state $\quasiIniSysDephased$. It would then be convenient to define the probability distribution  $\quasiPtpmGauss{\quasiw}{\quasiGaussianStdWT}$ that is the convolution of $ \tpmPwork(\quasiw)$ with a normal distribution with zero mean and variance $\quasiGaussianStdWT^2$
\begin{eqnarray}
  \quasiPtpmGauss{\quasiw}{\quasiGaussianStdWT}
  &=& \IntReals{u}  \tpmPwork(\quasiw-u) \NormalDist{0}{\quasiGaussianStdWT}{u} \label{eq:probconvoluted}\\
  &=& \sum_{n,m} \tr\left[ \quasiProjEfin_{m} \quasiUdriving
    \quasiProjEini_n \quasiIniSys \quasiProjEini_{n}
    \adj{\quasiUdriving} \right]
    \NormalDist{\quasiw_{nm}}{\quasiGaussianStdWT}{\quasiw}.
    \nonumber
\end{eqnarray}
Notice that $\quasiPtpmGauss{\quasiw}{\quasiGaussianStdWT}$ is simply the \TPM distribution, Eq.~\eqref{eq:def:wpovm:probwnm}, with the Dirac delta replaced by a normal distribution with the corresponding mean values of work and variance $\quasiGaussianStdWT^2$. Thus, for a highly localized normal distribution, it satisfies ${\quasiPtpmGauss{\quasiw}{\quasiGaussianStdWT} \xrightarrow[\quasiGaussianStdWT \to 0]{}  \tpmPwork(\quasiw)}$.

In order to illustrate the effect of initial coherences let us consider Eq.~\eqref{eq:def:wquasi:gaussian},
and split it in diagonal ($n = n'$) and non-diagonal ($n \neq n'$) contributions

\begin{equation}
  \quasiPwork(\quasiVarw, \quasiVart)
  = \quasiPtpmGauss{\quasiw}{\quasiGaussianStdWT}\NormalDist{0}{\frac{\hbar}{\sqrt 2\quasiGaussianStdWT}}{\quasiVart}
    +\quasiPwork^{\rm(c)}(\quasiVarw, \quasiVart).
  \label{eq:res:wquasi:gaussiandiagnondiag}
\end{equation}
The non-diagonal one corresponds to the contribution of  the so-called initial coherences and it is easy to see that
\begin{eqnarray}
\quasiPwork^{\rm(c)}(\quasiVarw, \quasiVart) &=& \sum_{n \neq n',m} \tr\left[ \quasiProjEfin_{m}
    \quasiUdriving \quasiProjEini_n \quasiIniSys(-\quasiVart) \quasiProjEini_{n'}
    \adj{\quasiUdriving} \right]  \\
    &&\times \ \NormalDist{\frac{\quasiw_{nm} + \quasiw_{n'm}}{2}}
      {\quasiGaussianStdWT}{\quasiVarw}
      \NormalDist{0}{\frac{\hbar}{\sqrt 2\quasiGaussianStdWT}}{\quasiVart}. \nonumber
\end{eqnarray}

\subsection{Quasidistribution for initial dephased states}

When the initial state of the system is diagonal in the energy basis ($\quasiProjEini_n \quasiIniSys
\quasiProjEini_{n'} = 0$ for $n \neq n'$), $\quasiPwork^{\rm(c)}(\quasiVarw, \quasiVart) =0$ and the Wigner function is just
\begin{equation}
  \quasiPwork(\quasiVarw, \quasiVart)
    = \quasiPtpmGauss{\quasiw}{\quasiGaussianStdWT}
      \NormalDist{0}{\frac{\hbar}{\sqrt 2\quasiGaussianStdWT}}{\quasiVart},
\end{equation}
that is, it's proportional to the convoluted \TPM distribution for every value of $\quasiVart$. Moreover, if we calculate the marginal $\quasiPwork(\quasiVarw)$,
\begin{equation}
  \quasiPwork(\quasiVarw)
  = \IntReals{\tau}\quasiPwork(\quasiVarw, \quasiVart)
    = \quasiPtpmGauss{\quasiw}{\quasiGaussianStdWT},
  \label{eq:res:wquasi:gaussiandiagwmarginal}
\end{equation}
we recover the probability distribution that would be obtained if one measures the observable $\wpovmAuxQop$. This expression just reflects a characteristic property of the Wigner function: the partial integration provides the probability distribution corresponding to the other variable.
Therefore, for initial states without coherences in the initial energy eigenbasis, $\quasiPwork(\quasiVarw)$ is exactly
$\quasiPtpmGauss{\quasiw}{\quasiGaussianStdWT}$. If in addition $\quasiGaussianStdWT 
 \ll (\quasiw_{nm} - \quasiw_{n'm'}), \forall
n,n',m,m'$ then we recover  the probability distribution of work given by the \TPM protocol.

\begin{figure*}[tb]
  \centering
  \includegraphics[width = \textwidth]{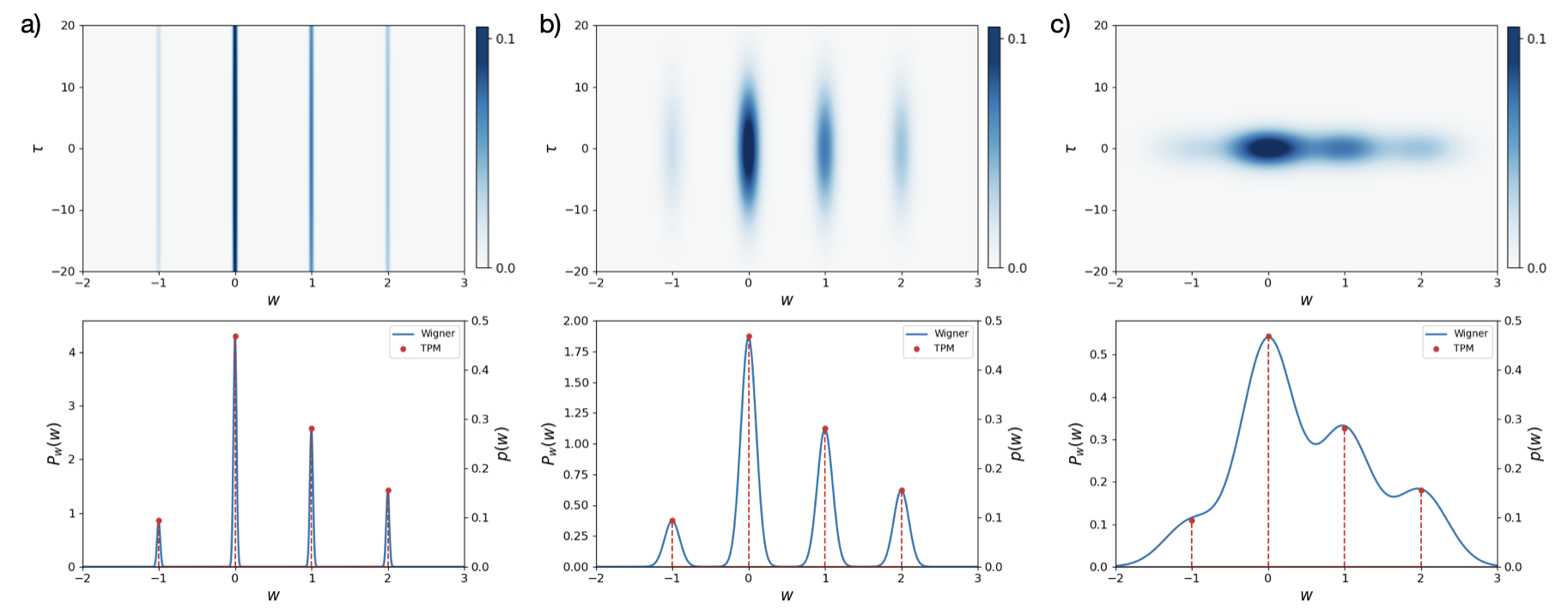}
  \caption{Wigner function of work for a two-level system using a Gaussian ancilla. The initial hamiltonian is $\quasiHini = \tpmEini\sigma_+\sigma_-$, with $\sigma_\pm$ the Pauli creation and annihilation operators.
    The unitary driving is given by $\tpmUdriving = (\sqrt{2}\Id +
    i\sigma_x + i\sigma_z)/2$ and the final hamiltonian is $\quasiHfin = 2\tpmEini\sigma_+\sigma_-$.
    The initial state of the system is $\quasiIniSys = (\Id +
    \sigma_z/4)/2$ and the variance of the initial Gaussian packets of the ancilla are (a) $\quasiGaussianStdWT = 0.02\tpmEini$, (b) $\quasiGaussianStdWT = 0.1\tpmEini$ and (c) $\quasiGaussianStdWT = 0.35\tpmEini$.
    The upper panel shows the distribution
    $\quasiPwork(\quasiVarw,\quasiVart)$ of 
    Eq.~\eqref{eq:res:wquasi:gaussiandiagnondiag} based on the Wigner function.
    In the lower panel we show the marginal of $\quasiVarw$, given by
    Eq.~\eqref{eq:res:wquasi:gaussiandiagwmarginal}, along with the discrete probabilities $p(\quasiVarw)$ corresponding to the usual \TPM distribution. 
    \label{fig:wquasi:gaussianincoh}}
\end{figure*}

In Fig.~\ref{fig:wquasi:gaussianincoh}(b), upper panel, we show the distribution $\quasiPwork(\quasiVarw,\quasiVart)$ for a two-level system $\quasiSys$
without initial coherences. In the lower panel, we show the marginal of the distribution in $\quasiVarw$,
$\quasiPwork(\quasiVarw)$, and compare it with the discrete probabilities associated to the \TPM.
Notice  that the area under each Gaussian in the marginal is equal to the corresponding probability in the \TPM protocol, and we can also see that it effectively reproduces the ideal \TPM distribution.
On the other hand, in Fig.~\ref{fig:wquasi:gaussianincoh}(a) we show the distribution of work obtained for the same system but using an ancilla that has an initial state with a standard deviation five times smaller. One can easily note that this case is much closer to the ideal projective measurement regime. In this case, the Wigner function is  invariant under translations in $\quasiVart$, as expected, since the initial state of the system commutes with the initial Hamiltonian. In Fig.~\ref{fig:wquasi:gaussianincoh}(c) we show the distribution for a standard deviation even bigger than the one in Fig.~\ref{fig:wquasi:gaussianincoh}(b). As we can see, while the position of the peaks matches the correct work values, there is a significant  overlap between the different Gaussians.

\subsection{Effects of quantum coherences}

Let us now consider a system with initial coherences. From Eq.~\eqref{eq:res:wquasi:gaussiandiagnondiag} we can notice that in this case the Wigner function also has Gaussian peaks on each work value $\quasiw_{nm}$, just as it happens for the dephased state.
However, there are some additional terms centered around the average of two work values with different initial energy,
$(\quasiw_{nm} + \quasiw_{n'm})/2$. These terms are the ones that hold the non-trivial dependence on the variable
$\quasiVart$ and, as we will see , they can be negative.
This can be easily seen from the following argument. If we look at Eq.~\eqref{eq:res:wquasi:gaussiandiagnondiag}, we have that
 $ \IntReals{\quasiVart}{\rm d\quasiVarw}\ \quasiPwork(\quasiVarw, \quasiVart)=1$, and, in addition, also the integral over the phase space of the first term is equal to one, as it is the Wigner function of the initial dephased state. Therefore, the integral of the second term must be zero. In order to do so,  the sum should attain some negative values since they are real. 
In these terms, except for the Gaussian modulation, the variable $\quasiVart$ only appears as a time evolution of the state.

The fact that time appears explicitly only for initial states with coherences has a clear interpretation. If the initial state $\quasiIniSys$ is diagonal, then it is a steady state of the initial Hamiltonian, and the 
state is the same for every instant in time before the driving.
On the other hand, if $\quasiIniSys$ has coherences, the state evolves due to the free evolution with the initial hamiltonian. 
This time, of course, is irrelevant at the moment of performing the first projecting energy measurement for the \TPM distribution. However, it appears in our approach due to the fact we are performing coherent operations between system and ancilla. Notably, one can also observe that the mean energy difference Eq.~\eqref{eq:cohEnergyDiff} is not invariant under initial time translations for states with coherences.
Thus, given a reference state $\quasiIniSys$, the calculated distribution contains, in principle, information about every initial state that is unitarily connected with $\quasiIniSys$ by the initial hamiltonian. Nevertheless, the amplitude of the Wigner function decays exponentially to zero when $\quasiVart\to\pm\infty$ due to the Gaussian modulation. 

At the same time, given the complementary nature of the variables work $\quasiVarw$ and time $\quasiVart$, when localizing the Gaussian in $\quasiVarw$ we are delocalizing it in $\quasiVart$. We will come back to this issue when we consider the marginals of the distribution.

\begin{figure*}[tb]
  \centering
  \includegraphics[width = \textwidth]{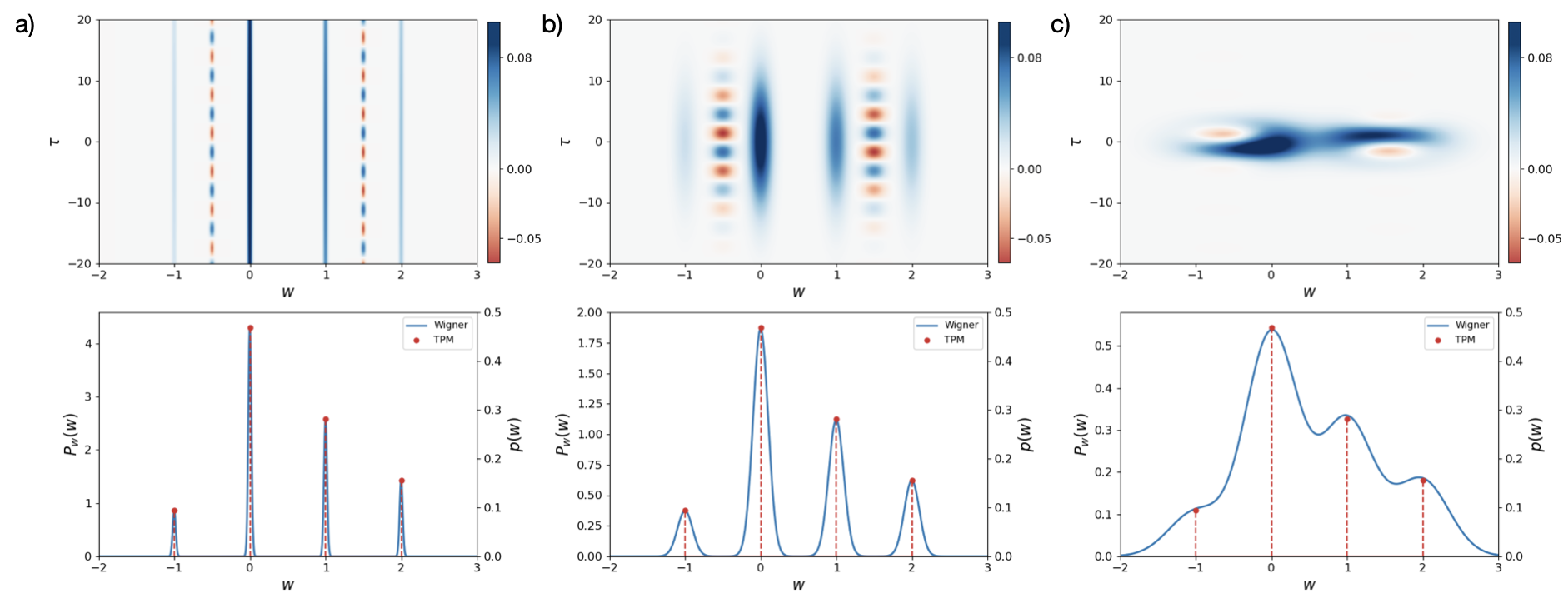}
  \caption{Wigner function for work of a two-level system using a Gaussian ancilla. The parameters used are the same as in Fig.~\ref{fig:wquasi:gaussianincoh}, except that now the initial density matrix has non-diagonal elements, $\quasiIniSys = (\Id
    + \sigma_x/2 + \sigma_y/2 + \sigma_z/4)/2$.
    The upper panel shows the distribution
    $\quasiPwork(\quasiVarw,\quasiVart)$
    Eq.~\eqref{eq:res:wquasi:gaussiandiagnondiag} based on the Wigner function. We notice now, because of of the initial  coherences, the appearence of negative values in the distribution. The lower panel shows the marginal of $\quasiVarw$, given by
    Eq.~\eqref{eq:res:wquasi:gaussiandiagwmarginal}, and it is compared to the work values and respective discrete probabilities $p(w)$ that appear in the usual \TPM distribution.
    \label{fig:wquasi:gaussiancoh}}
\end{figure*}

We have already shown 
that if the initial state does not have coherence in the energy basis, the resulting quasidistribution of work function is positive, because  the diagonal terms in Eq.~\eqref{eq:res:wquasi:gaussiandiagnondiag} are all positive. Therefore, if the distribution $\quasiPwork(\quasiVarw,\quasiVart)$ has some negativities, it is a signature of the presence of coherences in the initial state. This can be clearly seen in the upper panel of Fig.~\ref{fig:wquasi:gaussiancoh}, where the quasiprobability distribution of work $\quasiPwork(\quasiVarw,\quasiVart)$ of a two-level system is plotted. The hamiltonians, drivings and ancilla parameters are identical to those of Fig.~\ref{fig:wquasi:gaussianincoh}. 
Moreover, in both cases the initial state of the system has the same probability distribution in the energy basis.
The only difference between Figs.~\ref{fig:wquasi:gaussianincoh} and
\ref{fig:wquasi:gaussiancoh} is that in the second one the initial state has coherence between the two energy levels.
 Comparing both figures, we can notice that effectively we have the same Gaussian distributions over the same work values. The key difference lies in the fact that, for the initial coherent state, the quasidistribution displays additional oscillations that can become negative. This interference fringes indicate the presence of non-classicality in the Wigner function and in the initial state of the system.

In the lower panel of Fig.~\ref{fig:wquasi:gaussiancoh} we show  the marginal probability distribution for $\quasiVarw$.
Comparing them with Fig.~\ref{fig:wquasi:gaussianincoh} we can notice that the marginal distributions are equivalent.
In Fig.~\ref{fig:wquasi:gaussiancoh} we also show the work distribution for the same system but using an initial state of the ancilla $\quasiAux$ with different standard deviations.
In the case of the smaller standard deviation (corresponding to an ideal projective measurement) we can see that the marginal probability recovers that of the \TPM distribution. Notably, in the case of a bigger standard deviation the interference between different Gaussian picks modifies the distribution of the values of $\quasiVarw$, and there are corrections due to coherences as it is shown in Eq.~\eqref{eq:App:coherences}. In this limit,  the marginal distribution may not even coincide with that of the corresponding dephased state. This behaviour is similar to what happen when one makes a weak measurement~\cite{aharonov1988}. To understand when it is possible to observe these differences, lets note that when the marginal for $\quasiVarw$ is calculated from
Eq.~\eqref{eq:res:wquasi:gaussiandiagnondiag}, the diagonal terms give exactly the convoluted distribution $\quasiPtpmGauss{\quasiw}{\quasiGaussianStdWT}$.
For the non-diagonal contributions, we have time-averages of the form
\begin{equation}
  \IntReals{\quasiVart} \quasiIniSysTevol{-\quasiVart}
    \NormalDist{0}{\frac{\hbar}{\sqrt{2}\quasiGaussianStdWT}}{\quasiVart}.
    \label{eq:aver2}
\end{equation}
This operation is similar to a dephasing map in the energy basis, but there is a significant difference since this average is weighted with a normal distribution with variance $\hbar^2/(2\quasiGaussianStdWT^2)$ centered in the origin. The bigger the variance of the Gaussian (and therefore the smaller 
$\quasiGaussianStdWT$), more values of $\quasiVart$ enter in the time-average. Therefore, in the limit of small $\quasiGaussianStdWT$ we expect the non-diagonal terms to average to zero. Hence, one can show that if $\quasiGaussianStdWT  \ll (\quasiEini_{n} -
  \quasiEini_{n'})/2, \; \forall n,n'$, independent of the initial state, 
\begin{eqnarray}
  \quasiPwork(\quasiVarw)
  &=& \IntReals{\quasiVart}\quasiPwork(\quasiVarw,\quasiVart) \nonumber \\
  &\approx& \quasiPtpmGauss{\quasiw}{\quasiGaussianStdWT}.
\end{eqnarray}
Thus, the marginal of the quasiprobability distribution reproduces the \TPM distribution.

\subsection{Calculation of mean values}

Given the formalism associated to the Wigner function~\cite{hillery1984}, one can easily obtain average values from this quasidistribution. In fact, using the Wigner--Weyl representation~\cite{hillery1984} of an operator $\mathcal{A}$ acting on the ancillary space,
\begin{equation}
  A(\quasiVarw, \quasiVart)
  =  \IntReals{\quasiVarInt}
    \matrixel{\quasiVarw + \frac{\quasiVarInt}{2}}
      {\mathcal{A}}{\quasiVarw - \frac{\quasiVarInt}{2}}
    \ExpSuper{-i\quasiVart\quasiVarInt/ \hbar},
  \label{eq:def:wquasi:wwigner:operator}
\end{equation}
their mean value is just
\begin{equation}
    \tr\left[\mathcal{A}\ \quasiFinAux\right]=\PhaseSpaceInt \quasiPwork(\quasiVarw,\quasiVart)\ A(\quasiVarw, \quasiVart)
 \label{eq:def:wquasi:wwigner:meanvalues}.  
\end{equation}
For instance, the mean value of work is just the mean value of the operator $\wpovmAuxQop$, and it is obtained by integrating the function $w$ over the phase space
\begin{align}
   \mean{\quasiVarw} &\equiv \tr\left[\wpovmAuxQop\ \quasiFinAux\right]
   \nonumber \\
   &= \PhaseSpaceInt \quasiPwork(\quasiVarw,\quasiVart)\  w. 
\end{align}
The other typical average that is calculated in the context of fluctuation theorems, where the system is initially in thermal equilibrium at inverse temperature $\beta$, is $\mean{\ExpSuper{-\beta w}}$. This is easily done by integration of the function $\ExpSuper{-\beta w}$. In all cases,  the calculated mean values depend on the initial state of the ancilla. As it can be easily proven, for any observable of the type $f(\wpovmAuxQop)$, in the limit of $\sigma\rightarrow 0$ their averages converge to the values associated with the \TPM distribution.

\subsection{Energy difference in the presence of coherences}

Finally, we will show another interesting property that is contained in this quasidistribution. 
We have seen that, unless the initial state of the system is diagonal in the energy eigenbasis, the difference in mean energy and the mean value of work (Eqs.~\eqref{eq:cohEnergyDiff} and \eqref{eq:meanw}) do not coincide. Thus, the \TPM distribution does not provide any information about the initial coherences. Notably, as we will show, this information is also contained in this quasiprobability distribution. 

In order to do so, let us consider the average in phase space of the function
$g_{\quasiVart_0}(\quasiVarw,\quasiVart) =
\quasiVarw\ \DiracDelta(\quasiVart - \quasiVart_0)$ (see Appendix~\ref{sec:appQuasi}). Using the Wigner--Weyl transform~\cite{hillery1984}, it corresponds to the expectation value of the operator $\opmark{G}_{\wpovmAuxPeig_0} =
\left(\wpovmAuxQop\dyad{\quasiVart_0} +
\dyad{\quasiVart_0}\wpovmAuxQop\right)/2$ measured over the ancilla. It can be easily shown  that this average, which is equivalent to the integral of the function $\quasiVarw$ weighed by the Wigner function along an horizontal line at $\quasiVart_0$, is proportional to
\begin{equation}
   \PhaseSpaceInt \quasiPwork(\quasiVarw,\quasiVart)
    g_{\quasiVart_0}(\quasiVarw,\quasiVart)  \propto 
      \Delta E_{\quasiVart_0},  
\end{equation}
where $\Delta E_{\quasiVart_0}= \tr[ \quasiHfin \quasiUdriving \quasiIniSysTevol{-\quasiVart_0} \adj{\quasiUdriving}]- \tr[ \quasiHini\quasiIniSysTevol{-\quasiVart_0}]$ is the mean energy difference 
for a situation where the driving $\quasiUdriving$ is turned on at time $-\quasiVart_0$, and the proportionality constant is just equal to the Gaussian modulation at $\quasiVart_0$, $\NormalDist{0}{\frac{\hbar}{\sqrt 2\quasiGaussianStdWT}}{\quasiVart_0}$ (see Appendix~\ref{sec:appQuasi}). Therefore, when $\quasiVart_0=0$, this is just proportional to the `initial' energy difference $\Delta E$ in Eq.~\eqref{eq:cohEnergyDiff}. 
As we have shown, from this quasiprobability distribution we can calculate, not only the energy difference corresponding to the actual initial state, but 
also for the set of states $\quasiIniSysTevol{\quasiVart}$, $\tau \in \mathbb{R}$. This set can be viewed as different `initial times' at which the driving is turned on starting from a reference state $ \tpmRhoIni$ at time zero. This is so because this set of initial states is connected with $ \tpmRhoIni$ by a free hamiltonian evolution. 

Interestingly, for a Gaussian initial state of the ancilla, one obtains the correct value $\Delta E$ independent of their initial variance $\sigma$.
However, since there is a Gaussian modulation centered around $\quasiVart_0=0$ (the proportionality constant), the error in its determination increases as 
one localizes the initial state of the ancilla in the variable $w$. But, if one reduces the value of $\sigma$, the estimation of $P_\mathrm{\TPM}(\quasiVarw)$ gets worse. Therefore, one can also appreciate in this case the complementary nature of the variables $\quasiVarw$ and $\quasiVart$.


\section{Possible experimental implementations}
\label{sec:workdistribcoh:workquasi:experimental}
The measurement of the quasi-probability distribution that we propose requires two fundamental ingredients: (i) coherent control of two degrees of freedom of system and ancilla in order to implement the interactions of the \OPM protocol; (ii) being able to measure the Wigner function of the ancilla. In particular, implementing the \OPM requires the ability of performing translations of the ancilla conditioned on the energy degree of freedom on which the work is performed. There is a great variety of systems where this sort of interactions can be implemented, and an experimental realization of the \OPM protocol has been realised using cold atoms~\cite{cerisola2017}. However, it is not clear how one can implement the measurement of the Wigner function in such platform. Nevertheless, there are systems where both requirements are in principle satisfied and in what follows we will briefly describe two of them.

The first example is given by superconducting qubits coupled to a cavity, i.e. circuit quantum electrodynamics. In these systems, one has a charge qubit formed by a superconducting island coupled to a Josephson junction, and the two states correspond to the presence or absence of excess Cooper pairs in the island~\cite{makhlin2001}. Furthermore, the qubit circuit can be coupled to a wave-guide that acts as a microwave cavity where coherent states or states with a well defined number of photons can be stored~\cite{paik2011}.
For instance, in Ref.~\cite{naghiloo2018} they generate  coherent displacements of the state of the cavity depending on the state of the qubit. This interaction is exactly what is needed for implementing the protocol where the qubit acts as the system and the cavity as the ancilla. On the other hand, in a different coupling regime between qubit and cavity, this same scheme has been used to measure the Wigner function of the state of the field in the cavity~\cite{sun2014}.

The second possible platform  are trapped ions. In this case, ions are trapped in an electric potential such that the motional degrees of freedom of the ion are subjected to an effective harmonic oscillator potential~\cite{leibfried2003}. At the same time, using the interactions between the electronic degree of freedom and the position of the ion, it is possible to generate coherent, squeezed and Fock states of the oscillator~\cite{leibfried2003}. In particular, in different experiments~\cite{haljan2005,von-lindenfels2019} it has been shown that one can apply forces on the ion depending on its electronic state, and in this way displacements in phase space depending on the qubit state can be coherently implemented. Again, this is the interaction needed to perform the protocol. At last, the Wigner function of the motional degree of freedom of trapped ions has been successfully measured~\cite{leibfried1996}.

\section{Conclusions}
\label{section: conclusions}

In this work we introduced a generalization of the probability distribution of work based on the Wigner function. The starting point is the \OPM protocol proposed in~\cite{roncaglia2014}, where an ancilla is coupled to the system whose work one wants to measure in order to keep a coherent record of all possible work values. 
Following this idea, we define the Wigner function of the final state of the ancilla. 
This quasi-probability distribution contains all the information regarding both work and coherence in the initial state of the system. In fact, initial quantum coherence in the system results in negativities in the quasi-probability distribution of work, a clear signature of non classicality.  In this case, we can also recover the mean value of energy, which is different from the average work for states with coherences. Moreover, we show that from this quasi-probability distribution one can easily recover the standard \TPM distribution simply by integrating over the time variable. In addition, we show that given that the average work and other quantities of interest can be obtained as the mean value of an operator acting on the ancillary space, it is easy to calculate mean values using the formalism of the Wigner function.
The quasi-probability distribution here defined has certain similarities with the one proposed in~\cite{solinas2015,solinas2016,solinas2017}. The way in which the distribution is defined there, is also inspired on the \OPM scheme~\cite{de-chiara2018}, and requires the preparation of a coherent superposition of the ancilla between two momentum eigenstates, $\ket{p} +
\ket{-p}$, together with the implementation of an interaction analogous to that of the \OPM. At the end of the protocol, the relative phase between these states is measured and a quasiprobability distribution that contains information about work and coherence is obtained~\cite{de-chiara2018}. In contrast, our proposal has a clear operational interpretation and direct experimental application, as it is simply the Wigner function of the final state of the measurement apparatus. Moreover, our protocol not only contains all the information of Ref.~\cite{solinas2015,solinas2016,solinas2017}, but also for coherent initial states it has additional information on the dependence of the time variable, $\quasiVart$. From a practical point of view, our protocol does not need ideal (non-physical) states and it is easy to adapt to any initial state of the ancilla. Here, we have just developed the case of Gaussian states given that they are easy to treat analytically and are typically appropriate to model experimental conditions. However, this whole analysis can be repeated for any initial state.
We hope that this approach to the work distribution can shed some light to elucidate the effects of quantum coherences in thermodynamic transformations.

\begin{acknowledgments}
We thank C. Brukner and M. Saraceno for interesting discussions. 
This work was partially supported by CONICET, UBACyT, and ANPCyT. 
FC acknowledges support by grant number FQXi-IAF19-01 from the Foundational Questions Institute Fund, a donor advised fund of Silicon Valley Community Foundation.
\end{acknowledgments}

\appendix

\section{Quasi-probability distribution for Gaussian states}
\label{sec:appQuasi}
We start with the general expression in Eq.~\eqref{eq:def:wquasi:wwigner:general} for the Wigner function and replace the initial state of the ancilla with a coherent (squeezed) Gaussian state $\quasiIniAux = \dyad{\wpovmGaussianState{0}{\sigma}}$
centered in the origin of coordinates of the phase space and with variance $\quasiGaussianStd^2$
in $\wpovmAuxQop$ and $\frac{\hbar^2}{2 \sigma^2}$ in $\wpovmAuxPop$.
Then, we obtain
\begin{widetext}
\begin{align}
  \quasiPwork(\quasiVarw, \quasiVart)
  &= \frac{1}{2\pi\hbar} \sum_{n,n',m} \tr\left[ \quasiProjEfin_{m}
    \quasiUdriving \quasiProjEini_n \quasiIniSys \quasiProjEini_{n'}
    \adj{\quasiUdriving} \right] \times
  \nonumber \\
  &\qquad \times \frac{1}{\sqrt{2\pi}\quasiGaussianStd} \IntReals{\quasiVarInt} 
    \ExpInline{-\frac{\left(\quasiVarw + \frac{\quasiVarInt}{2} -
      \quasiw_{nm}\right)^2}{4\quasiGaussianStd^2}}
    \ExpInline{-\frac{\left(\quasiVarw - \frac{\quasiVarInt}{2} -
      \quasiw_{n'm}\right)^2}{4\quasiGaussianStd^2}}
    \ExpSuper{-i\quasiVart\quasiVarInt/\hbar}
  \nonumber \\
  &= \frac{1}{2\pi\hbar} \sum_{n,n',m} \tr\left[ \quasiProjEfin_{m}
    \quasiUdriving \quasiProjEini_n \quasiIniSys \quasiProjEini_{n'}
    \adj{\quasiUdriving} \right]
    \ExpInline{-\frac{\left(\quasiVarw - 
      \frac{\quasiw_{nm} + \quasiw_{n'm}} {2}\right)^2}{2\quasiGaussianStd^2}}
    \times
  \nonumber \\
  &\qquad \times \frac{1}{\sqrt{2\pi}\quasiGaussianStd} \IntReals{\quasiVarInt} 
    \ExpInline{-\frac{\left(\quasiVarInt -
      (\quasiw_{nm} - \quasiw_{n'm})\right)^2}{2\quasiGaussianStd^2}}
    \ExpSuper{-i\quasiVart\quasiVarInt/\hbar}
  \nonumber \\
  &= \sum_{n,n',m} \tr\left[ \quasiProjEfin_{m}
    \quasiUdriving \quasiProjEini_n \quasiIniSys \quasiProjEini_{n'}
    \adj{\quasiUdriving} \right]
    \NormalDist{\frac{\quasiw_{nm} + \quasiw_{n'm}}{2}}
      {\quasiGaussianStd}{\quasiVarw}
    \NormalDist{0}{\frac{\hbar}{\sqrt 2\quasiGaussianStd}}{\quasiVart}
    \ExpSuper{i
     \quasiVart (\quasiEini_{n} - \quasiEini_{n'})/\hbar},
  \label{eq:def:wquasi:wwigner:gaussian}
\end{align}

Now, let us calculate the marginal of Eq.~\eqref{eq:def:wquasi:wwigner:gaussian} in order to see that it gives us the convoluted probability distribution of work of Eq.~\eqref{eq:probconvoluted}. First, we split the function in diagonal and non-diagonal terms as in Eq.~\eqref{eq:res:wquasi:gaussiandiagnondiag}, and then integrate each of them:
\begin{align}
  \quasiPwork(\quasiVarw)
  &= \IntReals{\quasiVart}\quasiPwork(\quasiVarw,\quasiVart)
  \nonumber \\
  &= \quasiPtpmGauss{\quasiVarw}{\quasiGaussianStdWT}
    \IntReals{\quasiVart}
      \NormalDist{0}{ \frac{\hbar}{\sqrt 2\quasiGaussianStdWT}}{\quasiVart}
  \nonumber \\
  &\quad+\sum_{n \neq n',m} \tr\left[ \quasiProjEfin_{m}
    \quasiUdriving \quasiProjEini_n \quasiIniSys
    \quasiProjEini_{n'} \adj{\quasiUdriving} \right]
    \NormalDist{\frac{\quasiw_{nm} + \quasiw_{n'm}}{2}}
      {\quasiGaussianStdWT}{\quasiVarw}
    \IntReals{\quasiVart}
      \ExpSuper{i\quasiVart(\quasiEini_{n} - \quasiEini_{n'})/\hbar}
      \NormalDist{0}{\frac{\hbar}{\sqrt 2\quasiGaussianStdWT}}{\quasiVart}
  \nonumber \\
  &= \quasiPtpmGauss{\quasiVarw}{\quasiGaussianStdWT}
  +\sum_{n \neq n',m} \tr\left[ \quasiProjEfin_{m}
    \quasiUdriving \quasiProjEini_n \quasiIniSys
    \quasiProjEini_{n'} \adj{\quasiUdriving} \right] 
    \ \NormalDist{\frac{\quasiw_{nm} + \quasiw_{n'm}}{2}}
      {\quasiGaussianStdWT}{\quasiVarw}
    \ExpSuper{-\frac{(\quasiEini_{n} - \quasiEini_{n'})^2}
      {4\quasiGaussianStdWT^2}}.
      \label{eq:App:coherences}
\end{align}
Thus, if $\quasiGaussianStdWT \ll (\quasiEini_{n} -
\quasiEini_{n'})/2$, meaning that the dispersion is much smaller than all the energy gaps of the initial hamiltonian, then the non-diagonal terms become exponentially small and we have
\begin{equation}
  \quasiPwork(\quasiVarw)
  =\IntReals{\quasiVart}\quasiPwork(\quasiVarw,\quasiVart) 
  \approx  \quasiPtpmGauss{\quasiVarw}{\quasiGaussianStdWT}, \quad\quad \text{if}\; \quasiGaussianStdWT \ll \frac{\quasiEini_{n} -
  \quasiEini_{n'}}{2} \, \forall n,n'.
\end{equation}

Finally, let us consider the mean value of the 
$g_{\quasiVart_0}(\quasiVarw,\quasiVart) =
\quasiVarw\DiracDelta(\quasiVart - \quasiVart_0)$, this is equivalent to an average (weighted by the Wigner function) of 
work variable $\quasiVarw$ at a given fixed time $\quasiVart = \quasiVart_0$:
\begin{align}
   \PhaseSpaceInt \quasiPwork(\quasiVarw,\quasiVart) g_{\quasiVart_0}(\quasiVarw,\quasiVart) & = 
  \IntReals{\quasiVarw} \quasiPwork(\quasiVarw,\quasiVart_0) \quasiVarw  \nonumber \\
  &= 
    \NormalDist{0}{\frac{\hbar}{\sqrt 2\quasiGaussianStdWT}}{\quasiVart_0}
    \sum_{n,n',m} \tr\left[ \quasiProjEfin_{m}
    \quasiUdriving \quasiProjEini_n \quasiIniSysTevol{-\quasiVart_0}
    \quasiProjEini_{n'} \adj{\quasiUdriving} \right]
    \IntReals{\quasiVarw} \NormalDist{\frac{\quasiw_{nm} + \quasiw_{n'm}}{2}}
      {\quasiGaussianStdWT}{\quasiVarw} \quasiVarw
  \nonumber \\
  &=
    \NormalDist{0}{\frac{\hbar}{\sqrt 2\quasiGaussianStdWT}}{\quasiVart_0}
    \sum_{n,n',m} \tr\left[ \quasiProjEfin_{m}
    \quasiUdriving \quasiProjEini_n \quasiIniSysTevol{-\quasiVart_0}
    \quasiProjEini_{n'} \adj{\quasiUdriving} \right]
    \left(\frac{\quasiw_{nm} + \quasiw_{n'm}}{2}\right)
  \nonumber \\
  &= 
    \NormalDist{0}{\frac{\hbar}{\sqrt 2\quasiGaussianStdWT}}{\quasiVart_0}
    \sum_{n,n',m} \tr\left[ \quasiProjEfin_{m} \quasiUdriving \quasiProjEini_n
    \quasiIniSysTevol{-\quasiVart_0} \quasiProjEini_{n'} \adj{\quasiUdriving}
    \right]
    \frac{1}{2}\left(2\quasiEfin_m - \quasiEini_n - \quasiEini_{n'} \right)
  \nonumber \\
  &=
    \NormalDist{0}{\frac{\hbar}{\sqrt 2\quasiGaussianStdWT}}{\quasiVart_0}\left(
    \tr\left[ \quasiHfin\  \quasiUdriving \quasiIniSysTevol{-\quasiVart_0}
    \adj{\quasiUdriving} \right]-  \tr\left[ \quasiHini \quasiIniSysTevol{-\quasiVart_0} \right]
   \right)\nonumber \\
    &\equiv
    \NormalDist{0}{\frac{\hbar}{\sqrt 2\quasiGaussianStdWT}}{\quasiVart_0} \ \Delta E_{\quasiVart_0} \nonumber 
\end{align}
\end{widetext}
That is, this integral is proportional to the mean energy difference for an initial state, that may have coherences, when the driving is turned on at time $-\quasiVart_0$. Thus, for $\quasiVart_0=0$ it is the usual mean energy difference $\Delta E$ of \eqref{eq:cohEnergyDiff}.

\bibliographystyle{unsrt}
\bibliography{ref.bib}  
\end{document}